\documentclass[prl,aps,twocolumn,reprint,floatfix,showpacs,10pt]{revtex4-1}
\usepackage{epsfig}
\usepackage{graphicx}
\usepackage{epsfig}
\usepackage{color}
\usepackage{rotating}
\usepackage{amssymb}
\usepackage{amsmath}
\usepackage{bm}
\usepackage{graphics}
\usepackage{bm}
\usepackage{psfrag}
\usepackage{mathtools}
\usepackage[english]{babel}
\usepackage[T1]{fontenc}
\usepackage{color}
\usepackage{braket}
\usepackage[colorlinks,bookmarks=false,citecolor=blue,linkcolor=red,urlcolor=black]{hyperref}
\usepackage{epsfig}
\usepackage{umoline}

\def\beq{\begin{eqnarray}}
\def\eeq{\end{eqnarray}}

\newcommand{\be}{\begin{equation}}
\newcommand{\ee}{\end{equation}}
\newcommand{\bea}{\begin{eqnarray}}
\newcommand{\eea}{\end{eqnarray}}

\newcommand{\bk}{\mathbf k}

\begin{document}

\title{Different lattice geometries with synthetic dimension}
\author{Dominik Suszalski$^{1}$}
\email{dominik@suszalski.pl}
\author{Jakub Zakrzewski$^{1,2}$}
 \email{kuba@if.uj.edu.pl}

\affiliation{
\mbox{$^1$ Instytut Fizyki imienia Mariana Smoluchowskiego, Uniwersytet Jagiello{\'n}ski, ulica Reymonta 4, PL-30-059 Krak\'ow, Poland }
\mbox{$^2$ Mark Kac Complex Systems Research Center,
Uniwersytet Jagiello\'nski, Krak\'ow, Poland }
}

\begin{abstract}
The possibility of creating different geometries with the help of an extra synthetic dimension in optical lattices is studied. Additional linear potential and Raman assisted tunnelings are used to engineer well controlled tunnelings between available states. The great flexibility of the system allows us to obtain different geometries of synthetic lattices with possibility of adding synthetic gauge fields.
\end{abstract}


\maketitle

Cold atoms in optical lattices are one of prospective tools for quantum simulations in the spirit originally anticipated by Feynman \cite{Feynman82,Feynman86}. Various ingenoeus schemes have been proposed (for some recent reviews see \cite{Lewenstein12}). The
great flexibility of these systems results from controllability of optical lattice parameters as well as possibility of
using the internal level structure of atoms held in optical potential. This resulted in simulation of different models ranging from magnetism \cite{Eckardt10,Struck11} to abelian and non-abelian gauge theories
\cite{Zohar11, Banerjee12, Zohar12, Zohar13,Tagliacozzo13, Luca13,Zohar13a,Kosior14}.

An interesting observation has been made recently \cite{Boada12} that the structue of internal atomic states (e.g. magnetic sublevels) may be treated as an extra dimension. Thus for the real system of dimension $D$ a model of $D+1$ dimensions may be build. Interestingly the same group was able to show that synthetic gauge fields may be created in synthetic dimension \cite{Celi14}. It has been pointed out that sharp edges (open boundary conditions) realized naturally in the synthetic dimension may help in observation of chiral edge states as verified experimentially not much later \cite{Stuhl15,Mancini15}. The proposition \cite{Celi14} has stimulated interesting new ideas ranging from suggestion to realize Hall physics in $D=4$ \cite{Price15},
simulating Weyl semimetal physics \cite{Zhang15}, creating baryon squishing \cite{Ghosh15} or charge pumping effects \cite{Zeng15}.  

In all these application the internal degrees of freedom played the rule of an extra dimension separated from
the ``real'' dimensions. The aim of this letter is to expand the synthetic dimension concept to the case when real and synthetic dimensions become entangled (or not separable). As examples we show that such an approach allows us to create triangular or hexagonal lattices.

We consider a simple quasi one-dimensional system with the optical lattice along the $x$ direction. In the transverse directions we assume a tight confinement limiting the real space dynamics to one dimension for simplicity. The extension of
the proposed scheme to more dimensions is straightforward. 
Typically the hopping in the synthetic dimension is provided via Raman coupling, while hopping in the real dimension is the usual kinetic tunneling  term \cite{Boada12,Celi14,Price15,Zhang15,Ghosh15,Zeng15}. In our model we propose to utilize, in addition to an optical lattice, an additional linear potential, which effectively creates Wannier-Stark Hamiltonian \cite{Gluck02, Ketterle,Miyake13,Mishra15}:

\begin{equation} \label{ham}
H_{WS} = \frac{\hat{p}^2}{2m} + V \cos^2(k_L \hat{x}) + F \hat{x},
\end{equation}
(we shall adopt the recoil energy $E_R=\hbar^2 k_L^2/2,$ as the energy unit with unit of length given by the lattice constant $a=\pi/k_L$ where $k_L$ is the wavevector of the standing wave pattern forming the optical lattice. We set  $a=1$).

In the tight binding approximation the lowest band of this Hamiltonian is given by Wannier-Stark ladder \cite{Gluck02} with wavefunction referred to as  Wannier-Stark functions. They are linear combinations of Wannier functions of the original optical lattice. This picture is valid when the energy offset between two neighboring sites is bigger than the kinetic tunneling in the absence of a linear potential and smaller than the energy gap to the excited bands (where Zenner tunneling additionally appears)  \cite{Gluck02}. Within this approximation the standard nearest neighbour  tunnelings are included by passing to the Wannier-Stark basis - compare Appendix. The 
 desired tunnelings   may be  induced via  resonant Raman processes \cite{Ketterle, Miyake13,Greschner15} with a  great freedom.
One could also use a constant magnetic field splitting magnetic sublevels - creating an additional tilt in the synthetic direction. 

In the presence of the linear potential the Raman process is due to  two laser beams with frequencies tuned to exactly compensate the energy offset between connected sites. That induces controllable, possibly complex tunnellings in the system. Denote by $\mathbf q$ the momentum transfer  occuring during the Raman process, while by $q$ its projection on the $Ox$- axis. Then the Raman induced tunneling is expressed by \cite{Miyake13} (see also supplementary material in \cite{Greschner15}):

\begin{equation}
t_{j,k}=\Omega_{\mathrm eff} \int W^S_j (x) W^S_k (x) e^{-i q x} dx,
\label{rate}
\end{equation}
where $W^S_i(x)$ is a Wannier-Stark function at the (physical) site $i$ while $\Omega_{eff}$ is the effective strength of the Raman process.
For Raman lasers detuned by $\delta$  from a single upper level $\Omega_{\mathrm eff}= \Omega_1 \Omega_2/4\delta$ with $\Omega_1$ and $\Omega_2$ being  one-photon Rabi frequencies of the couplings. Often
one should consider simultaneous couplings with the entire hyperfine manifold, that modifies $\Omega_{eff}$  leaving the integral in \eqref{rate} unaffected \cite{Goldman14}). 
\begin{figure}
\includegraphics[width=85 mm]{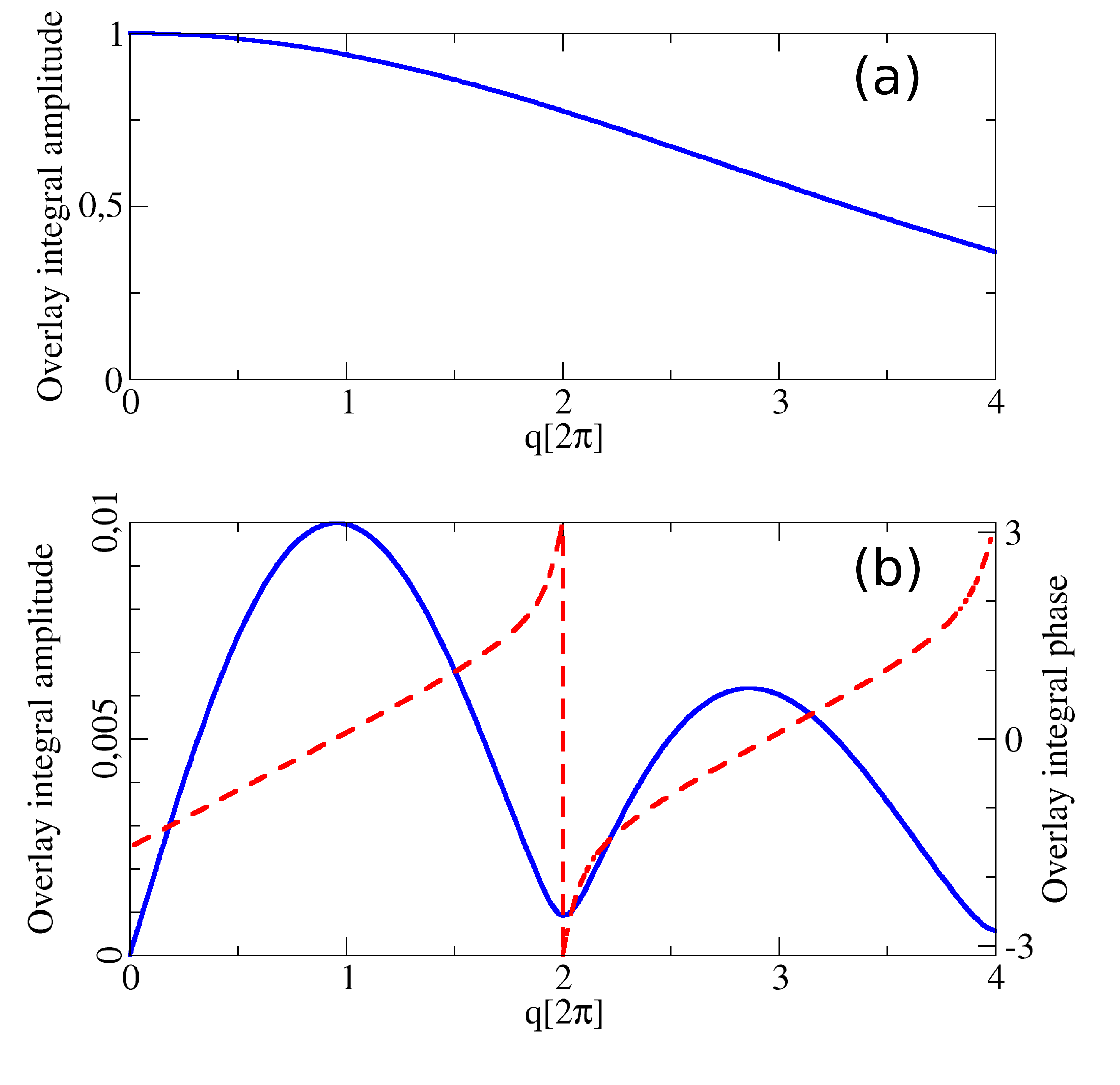}
\caption{The integral in Eq.~\eqref{rate} for $j=k$ (a) and $j=k+1$ (b). $V=20E_R$ and $F=0.15$ in that lattice depth the bare tunneling between wannier functions $t\approx-2.5\cdot10^{-3}$. Both the modulus (amplitude - blue) and the phase (red) of the integrals are plotted. Observe that the amplitude of the transition for Raman process involving the change of the physical site (b) is orders of magnitude smaller than those for the same physical site (a). To have similar amplitudes for those processes laser intensities and detunings that determine different $\Omega_{\mathrm eff}$ factors may be adjusted appropriately.  }
\label{fig:int}
\end{figure}

There are several possibilities. Consider a Raman process that couples two Zeeman sublevels of atoms residing in the same physical
(along $Ox$) site.  Taking the Zeeman sublevel quantum number $m$ as a site index in the synthetic dimension, by appropriately chosing the polarizations of Raman lasers one can realize both nearest-neighbor tunneling (with $\Delta m=\pm 1$) as well as
next nearest-neighbor tunnelings (with $\Delta m=\pm 2$). Then $j=k$ in \eqref{rate} and the tunneling rate reduces to
\be\label{rate0}
t_j=\Omega_{\mathrm eff} e^{-i q_s j}\int |W^S_0(x)|^2 e^{-i q_s x}dx, \ee
 with Wannier-Stark function $W_0(x)$ centered around 0, $q_s$ being the $x$ component of the momentum transfer in Raman process and $j$ numbering sites along the physical direction. The tunneling in the synthetic direction is thus complex with the phase (called later the spacial phase)  being site-dependent \cite{Celi14}. The integral contributes a real amplitude only as $|W^S_0(x)|^2$ is symmetric around the origin  [compare Fig.~\ref{fig:int}(a)]. This synthetic tunneling cobined with standard tunneling along $Ox$ (without the electric field tilt)
allows \cite{Celi14} for creating gauge fields and edge states. The flux through the square plaquete is then equal to $\Phi_S=-q_s$
and may be adjusted changing the transfer $q_s$ (subscript indicating the synthetic direction). 

In the presence of the static electric field, in the Wannier-Stark basis, the standard kinetic tunnelings are nonresonant (and included in the basis transformation). The desired coupling can be induced by another Raman assisted tunnelings along the real physical axis of the system and is given by
\begin{equation}
t_{j,j+1}=\Omega_{\mathrm eff} e^{-i q_x j}\int W^S_0(x) W^S_1 (x) e^{-i q_x x} dx,
\label{rate1}
\end{equation}
resulting in two phase factors. The first, as before, is the spatial phase and stands in front of the integral. The second, 
 the phase of the integral present in the formula above (called the overlap integral) will be referred to as  the phase of the overlap integral  [and is depicted in Fig.~\ref{fig:int}(b)]. The sum of both phases gives the phase of tunneling amplitude. Combined with the synthetic tunneling described above (generated by another pair of laser beams) we get another realization of the Harper Hamiltonian resembling that of \cite{Celi14} and involving a real and a synthetic dimension. The emerging lattice in this scheme is topologically equivalent to square lattice.
 
The next possibility is realized by Raman assisted tunneling involving both  the jump along the real and the synthetic directions.
We call this a diagonal process.
That process is also described by \eqref{rate1}, for the sake of clarity we shall denote the corresponding wavevector transfer $q_x^s$.

The flexibility of different Raman coupling lattice schemes we show on several examples. We assume, for simplicity the case $J=1$ in the ground state leading to $2J+1$ sublevels. Most of the schemes generalize to arbitrary $J$, one has to take into account the fact that the tunneling amplitudes become then site dependent (along synthetic dimension) as modified by appropriate Clebsh-Gordan coefficients modifying Rabi frequencies in  
\eqref{rate}.

\begin{figure}
\includegraphics[width=80mm]{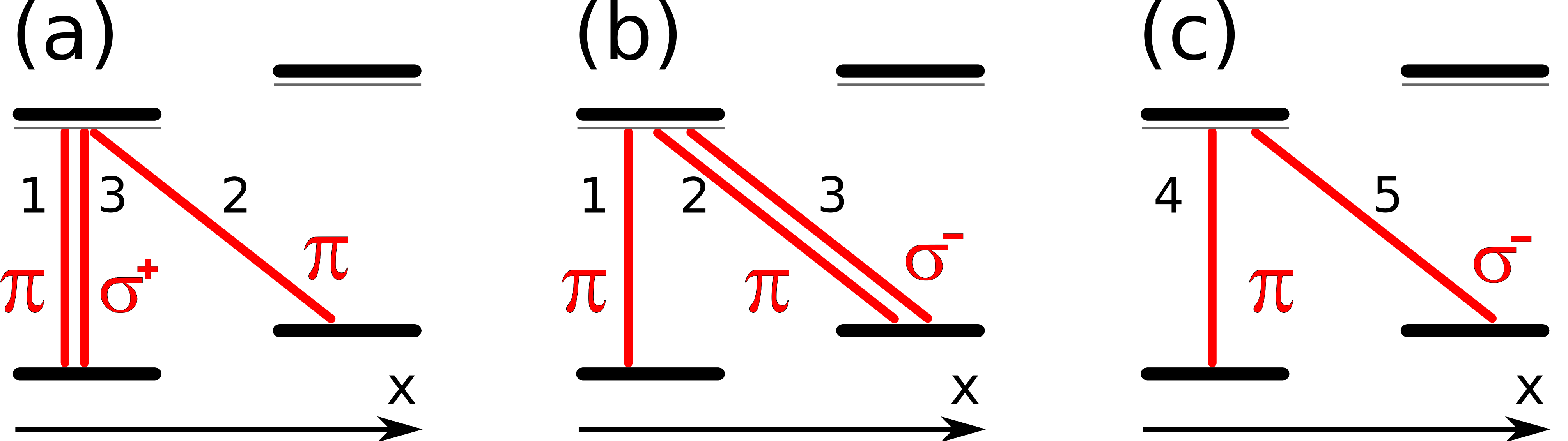}
\caption{(a) The configuration of laser beams used to create triangular lattice. The tunneling in synthetic direction is generated by beams 1 and 3, in the spatial direction by 1 and 2, while in the diagonal direction by 2 and 3 (see Fig.~\ref{lattice-triangular}). (b) The configuration of lasers used to create Harper Hamiltonian. The tunneling in spatial direction is generated by beams 1 and 2, while in the diagonal direction by 1 and 3 (see Fig.~\ref{Harper}). c) The configuration of laser beams used to create second of two lattices used to generation of hexagonal lattice.}
\label{figure-lasers}
\end{figure}

\begin{figure}
\includegraphics[width=80mm]{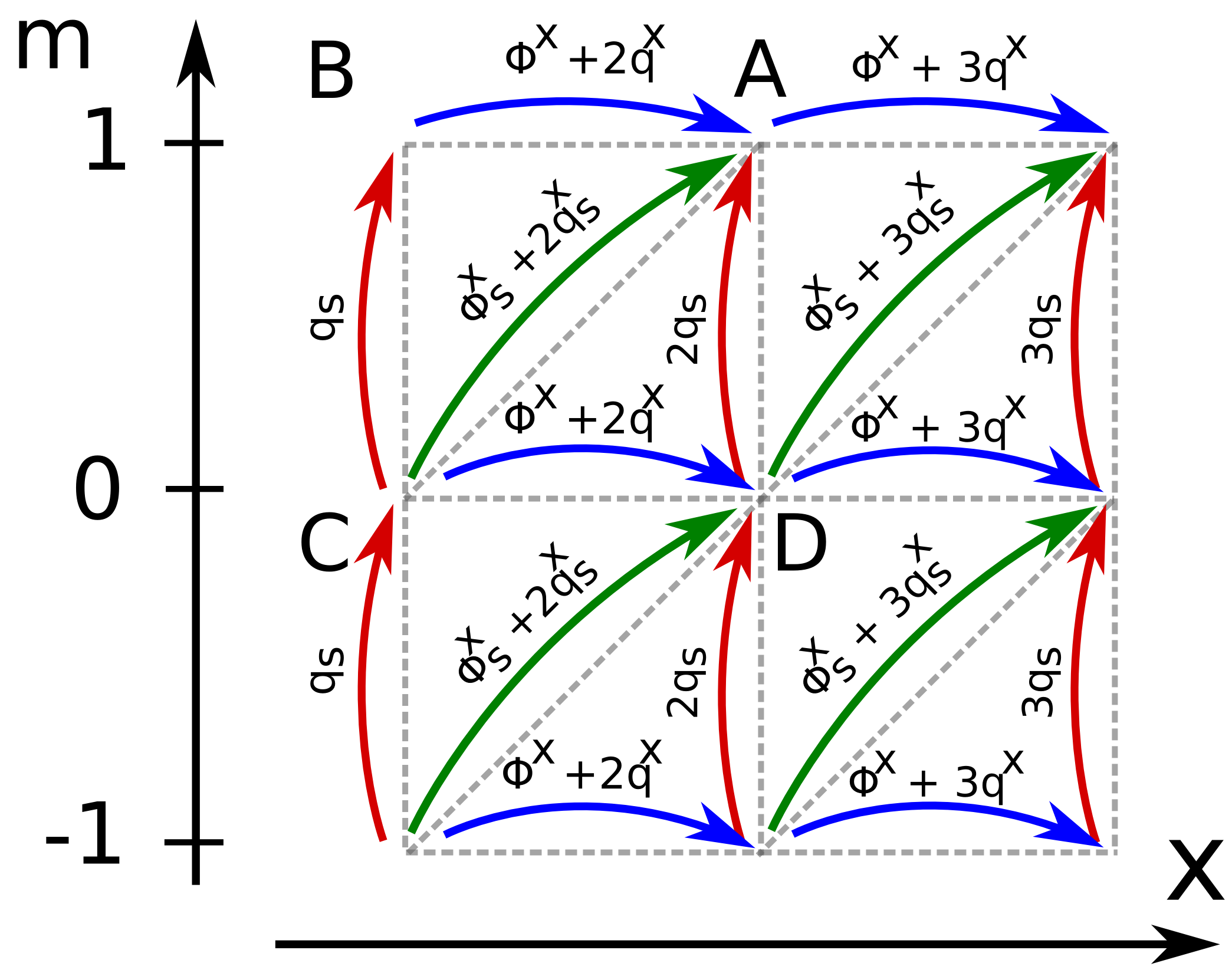}
\caption{The triangular lattice with additional magnetic field. The arrows represent tunneling between sites. The magnetic flux (in standard units) through the plaquette ABC is equal to $q_s^x - q^x  + \phi_s^x - \phi^x$, while magnetic flux trough DAC plaquette is equal to $ q_s - q_s^x + q^x - \phi_s^x + \phi^x$.}
\label{lattice-triangular}
\end{figure}

\textit{Emerging triangular lattice.} We shall involve all three Raman processes described above introducing three laser beams which couple neighboring states in a way depicted in  Fig.~\ref{figure-lasers}(a). The corresponding lattice is shown in the Fig.~\ref{lattice-triangular}. Laser beams 1 and 2,
both of $\pi$ polarization, generate tunneling along the physical dimension with $q_x=(\bk_1-\bk_2)\mathbf{\hat x}$, where $\bk_i$ - wavevector of the $i$-th lase beam. The same laser beam 1 with the another beam 3 (of $\sigma_+$ polarization) may create tunneling along the synthetic direction, while a pair of beams 3 and 2
may induce tunneling in both synthetic and real directions. All Raman processes may be made resonant simultaneously. Each laser beam is involved in the generation of two different tunneling processes, that allows to create three tunnelings with only three beams, while usually each Raman process requires two beams.

In the absence of the diagonal tunneling (for zero momentum transfer $q^s_x=(\bk_3-\bk_2)\hat x$) we obtain Harper-like Hamiltonian \cite{Celi14} with flux $\Phi=q_s$ through the plaquette ABCD (see Fig.~\ref{lattice-triangular}). The diagonal tunneling adds additional staggered fluxes across triangular plaquettes ABC and BDC with the sum of the two being $\Phi$. The amplitude of the tunnelings in different directions depends on the integrals of Wannier-Stark functions but can be adjusted via modifications of Raman laser intensities and detunings.

\begin{figure}
\includegraphics[width=80mm]{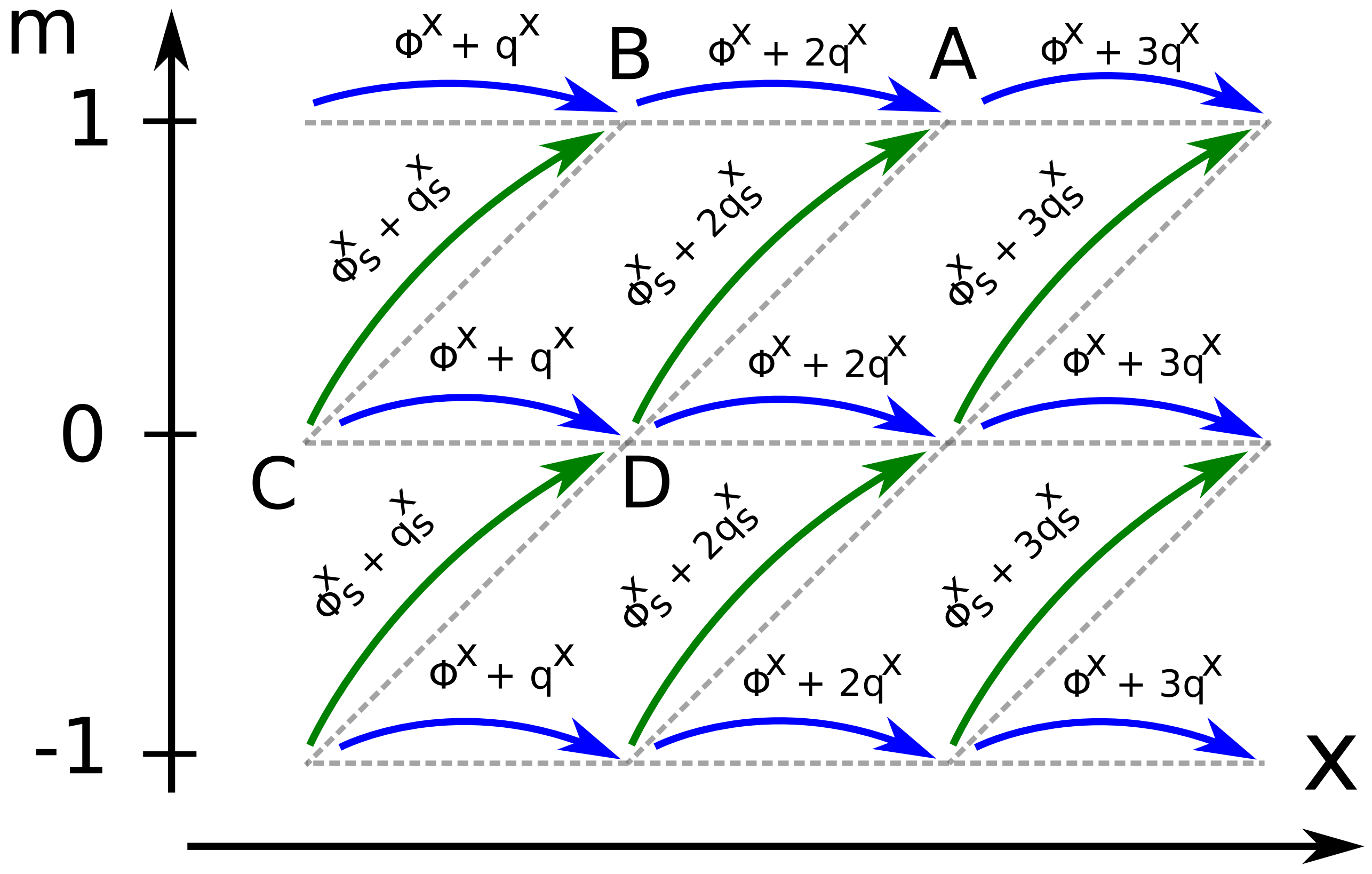}
\caption{The square lattice with additional magnetic field - realization of Harper Hamiltonian in synthetic dimension. The arrows represent tunneling between sites. The magnetic flux (in standard units) through the plaquette ABCD is equal to $q^x - q_s^x$.}
\label{Harper}
\end{figure}

\textit{The Harper Hamiltonian.}
The Harper Hamiltonian, the paradigm Hamiltonian of magnetism can also be obtained in the present scheme in a different way 
(note earlier realizations of it  in 2D systems \cite{Aidelsburger11,Miyake13}). As before, three laser beams are needed (see Fig.~\ref{figure-lasers} (b)) and the resulting geometry of the lattice is shown in  Fig.~\ref{Harper}).  The  lattice is tilted in comparison to other schemes involving the synthetic dimension that have been proposed till now \cite{Celi14,Barbarino15} though the topology is preserved. The magnetic flux is easily tunable by the difference of spatial phases of $q^s_x-q_x=(\bk_2-\bk_3)\hat{x}$ and the overlap integral phase is an additional minor modification (which does not influence the dynamic of the system). Formally there exist a process generated by beams 2 and 3 which would be a hopping term along synthetic dimension, thought there are two ways of removing it. The first possibility is to set $q^s = (\bk_2-\bk_3)\hat{x} = 0$ (just like it is described in the case of emerging triangular lattice), but it will remove the synthetic magnetic field from our system. The second involves 
the fact, that in that proces the overlap integral present in the formula 
\ref{rate1} appears twice making it two orders of magnitude less intensive, than desired tunnelings. One could improve that ratio, by modyfing the relative intensities of laser beams gaining additional two orders of magnitude. While  the triangular lattice construction described above depends more heavily on the overlap integral phase factor, for the present realisation
of the Harper Hamiltonian it plays no role.  In the construction of emerging honeycomb lattice (see below)  the overlap integral phase factor plays again an important role. 

\begin{figure}
\includegraphics[width=80mm]{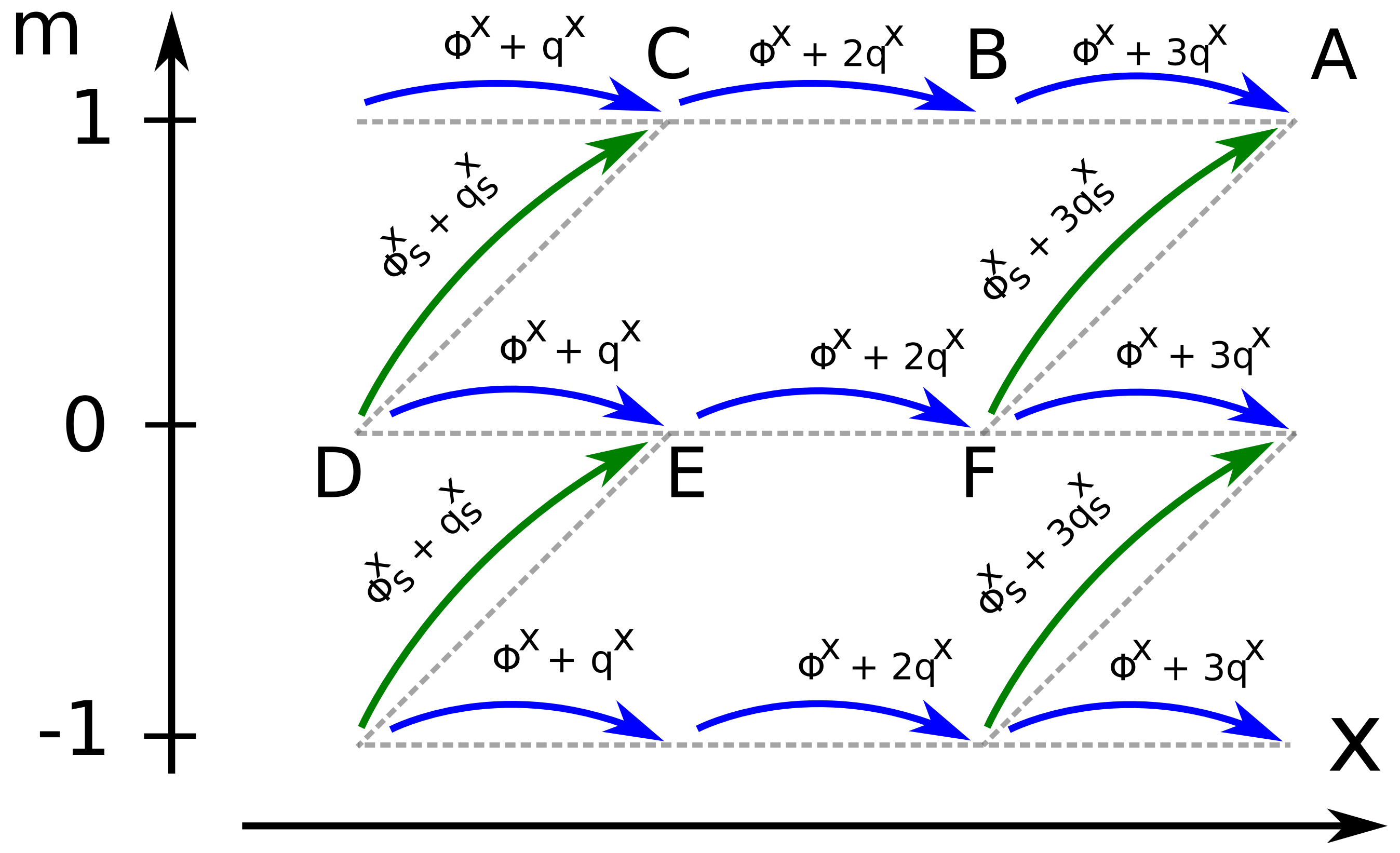}
\caption{The artificial hexagonal lattice in synthetic dimension. The flux of magnetic field through the ABCDEF plaquette is equal to $2(q^x-q_s^x)$.}
\label{lattice-fullhex}
\end{figure}

\textit{The emerging honeycomb lattice} 
In order to create the emerging honeycomb lattice one can modify the Harper Hamiltonian proposition [see Fig.~\ref{Harper}]  and
add an additional pair of laser beams [Fig.~\ref{figure-lasers}(c)] creating yet another Raman process. The second pair  of laser beams creates only the diagonal tunneling (changing both the real and the synthetic sites) and the emerging lattice results form the coherent superposition of two Raman processes, the one applied in the Harper Hamiltonian proposition and this additional one. Using the flexibility in adjusting tunneling phases one may realize the situation when the diagonal tunneling amplitudes generated by both Raman processes
add  constructively (destructively) on consequtive even (odd) diagonals. When both Raman processes involved have the same amplitude, every second diagonal tunneling will cancel out due to the destructive interference leading to the brick-like lattice structure, topologically equivalent to the honeycomb lattice - compare Fig.~\ref{lattice-fullhex}. In such a realisation the magnetic flux through the plaquette ABCDEF (see Fig.~\ref{lattice-fullhex}) is given by $2(\textbf{k}_2-\textbf{k}_3)\hat{x}$ so it is possible to create any magnetic flux through such a plaquette (the maximal magnetic flux is $2\pi$).

In order to obtain such a cancelation one has to ensure that the spatial phase of the second tunneling is greater than the spatial phase of first tunneling by $\pi$. Moreover, one has to control the relative phase  between laser fields involved in both Raman processes. As these lasers are assumed to work at significantly different frequencies they may be made coherently coupled via locking with the frequency comb set-up. The convenient arrangement could be when this relative phase cancels the difference of phases of overlap integrals in the diagonal tunnelings. The additional requirement of the equality of amplitudes may be satisfied by modification of laser intensities provided that both spin structure and hyperfine manifold of the excited levels used in Raman processes are identical. Let us note that such a double-Raman scheme is in fact a small modification of the standard double-lambda optical scheme (albeit applied in optical lattice setting)  used for a long time in laser spectroscopy (see e.g. \cite{
Zanon05} and references therein) and the phase control necessary for our scheme is similar to that required for laser spectroscopy coherent effects.

\textit{Conclusions} 
We have shown that a combination of Raman transitions in tilted 1D optical lattice allows to create lattices of different geometries mixing the real and the synthetic dimension. In particular the possibility to generate an emergent triangular as well as honeycomb
lattices  is discussed. 

After our work was completed
Gediminas Juzeli\=unas was kind enough to share with us the preliminary results  of another study of
nonseparable lattice involving a synthetic dimension in the special case of zig-zag ladder including the
interaction analysis \cite{topsecret}. Discussions with him as well as wi
 with Jacopo Catani, Leonardo Fallani, Wojtek Gawlik, and  Luis Santos at various stages of this work are acknowledged. We thank Krzysztof Biedro\'n for sharing his program for calculating Wannier functions. this research was performed within 
project   No. DEC-2012/04/A/ST2/00088 financed by  National Science Center (Poland). We also acnowledge support of EU via project QUIC (H2020-FETPROACT-2014 No.641122). 

\bibliography{synthetic}

\appendix

\section{Supplementary material: Wannier-Stark problem}

We want to find solutions of the Wannier-Stark Hamiltonian:

\begin{equation}\label{aa}
H_{WS} = H_0 + F \hat{x},
\end{equation}

where:

\begin{equation}
H_0 =  \frac{\hat{p}^2}{2m} + V \cos^2(k_L \hat{x}).
\end{equation}

The standard solutions of $H_0$ are Bloch functionslabelled by quasimomentum $k$ and the band number. for low temperatures  the higher bands may be omitted due to a finite energy gap between bands. From the Bloch functions one may construct the basis of exponentially localized states - the Wannier functions given by the formula:

\begin{equation}
\psi_l(x) = \int_{-\pi/2}^{\pi/a} \exp(-i k l a) \phi_k(x) dk,
\end{equation}
where $\psi_l(x)$ is the Wannier function localized in l-th site, $a=\frac{\pi}{k_L}$ is lattice constant and $\phi_k(x)$ is Bloch function from the lowest band with quasimomentum $k$. Wannier functions satisfy relation $\psi_{l+1}(x) = \psi_l(x-a)$, which one would expect from translational symmetry. The Wannier functions are exponentially localized. In second quantization formalism and keeping only the nearest neighbour tunnelings, the standard tight-binding representation of $H_0$ reads

\begin{equation}
H_0 \approx \sum_{l} \epsilon a^\dagger_{l}a_{l} - \sum_{l} {t a^\dagger_{l+1} a_{l} + h.c.}.
\end{equation}

Adding now the linear tilt as in \eqref{aa} and incorporating only the diagonal part of the linear potential into the tight biding model results in the Hamiltonian:

\begin{equation}
H_{TB} = \sum_{l} (\epsilon + aFl) a^\dagger_{l}a_{l} - \sum_{l} {t a^\dagger_{l+1} a_{l} + h.c.}.
\end{equation}

The exact solution of that hamiltonian are Wannier-Stark functions, defined by:

\begin{equation}
\omega_l(x) = \sum_{m} J_{m-l}\left( \frac{2t}{aF} \right) \psi_m (x).
\end{equation}

For more information see \cite{Gluck02}.

\end{document}